\documentclass[preprint,12pt]{elsarticle}

\usepackage{amssymb}
\usepackage{subcaption}
\usepackage{array}
 \usepackage{multirow}
\usepackage{hyperref}   
\usepackage{booktabs}
\usepackage{amsmath}
\usepackage[utf8]{inputenc}

\usepackage{graphicx} 
\usepackage{multirow} 
\usepackage{hyperref} 
\usepackage[utf8]{inputenc} 
\usepackage[T1]{fontenc} 

\usepackage[pdf]{graphviz}
\usepackage{xpatch}
\usepackage{graphicx}
\usepackage[table,xcdraw]{xcolor}
\makeatletter
\newcommand*{\addFileDependency}[1]{
  \typeout{(#1)}
  \@addtofilelist{#1}
  \IfFileExists{#1}{}{\typeout{No file #1.}}
}
\makeatother
\xpretocmd{\digraph}{\addFileDependency{#2.dot}}{}{}

\begin{document}

\begin{frontmatter}

\title{Defining Effective Engagement For Enhancing Cancer Patients'  Well-being with Mobile Digital Behavior Change Interventions}

\author[inst1,inst2]{Aneta Lisowska}

\affiliation[inst1]{
organization={Department of Computer Science},
addressline={Vrije Univerisiteit Amsterdam},
city={Amsterdam},
country={Netherlands}}

\affiliation[inst2]{
organization={Cancer Center Amsterdam},
addressline={VU University Medical Center (VUmc)},
city={Amsterdam},
country={Netherlands}}

\author[inst3]{Szymon Wilk}

\affiliation[inst3]{organization={ Institute of Computing Science},
            addressline={Poznań University of Technology}, 
            city={Poznan},
            country={Poland}}

\author[inst4,inst5]{Laura Locati}

\affiliation[inst4]{organization={Department of Internal Medicine and Medical Therapy},
            addressline={University of Pavia},
            city={Pavia},
            country={Italy}}
\affiliation[inst5]{organization={Medical Oncology Unit}, addressline={Istituti Clinici Scientifici Maugeri IRCCS},
            city={Pavia},
            country={Italy}}

\author[inst6]{Mimma Rizzo}

\affiliation[inst6]{organization={Division of Medical Oncology},
            addressline={Azienda Ospedaliero Universitaria Consorziale Policlinico di Barii},
            city={Bari},
            country={Italy}}

\author[inst7]{Lucia Sacchi}
\author[inst7]{Silvana Quaglini}
\affiliation[inst7]{organization={Department of Electric, Computer and Biomedical Engineering},
            addressline={University of Pavia}, 
            city={Pavia},
            country={Italy}}

\author[inst8]{Matteo Terzaghi}
\author[inst8]{Valentina Tibollo}

\affiliation[inst8]{organization={Istituti Clinici Scientifici Maugeri IRCCS},
            city={Pavia},
            country={Italy}}

\author[inst9]{Mor Peleg}

\affiliation[inst9]{organization={Department of Information Systems},
            addressline={University of Haifa}, 
            city={Haifa},
            country={Israel}}

\begin{abstract}

Digital Behavior Change Interventions (DBCIs) are supporting development of new health behaviors. Evaluating their effectiveness is crucial for their improvement and understanding of success factors. However, comprehensive guidance for developers, particularly in small-scale studies with ethical constraints, is limited. Building on the CAPABLE project, this study aims to define effective engagement with DBCIs for supporting cancer patients in enhancing their quality of life. We identify metrics for measuring engagement, explore the interest of both patients and clinicians in DBCIs, and propose hypotheses for assessing the impact of DBCIs in such contexts. Our findings suggest that clinician prescriptions significantly increase sustained engagement with mobile DBCIs. In addition, while one weekly engagement with a DBCI is sufficient to maintain well-being, transitioning from extrinsic to intrinsic motivation may require a higher level of engagement.

\end{abstract}

\begin{keyword}
digital behavior change interventions \sep effective engagement \sep cancer \sep well-being \sep evaluation

\end{keyword}

\end{frontmatter}


\section{Introduction}

Digital behavior change interventions (DBCIs) are a promising tool for assisting patients in creating new health habits that could positively impact their well-being. Evaluating the effectiveness of DBCIs is a crucial step in their development life cycle \cite{mummah2016ideas} and in allowing the research community to learn which properties of DBCIs and their implementation strategies make them successful. Several papers discuss the challenges inherent in evaluating DBCI's effectiveness and suggest recommendations for evaluation methods \cite{yardley2016understanding,michie2017developing}. Identifying effective engagement with DBCIs in specific contexts has become a critical aspect of this discourse. 

However, the state of the art does not provide detailed guidance, which leaves developers of DBCIs in need of further instruction. The problem is amplified for small-scale evaluation studies that do not allow for large groups of patients and in some cases do not include proper control groups, due to ethics considerations. Hence, the goal of this study is to describe a sequence of hypotheses, incorporating the concept of effective engagement as a critical factor, that should be evaluated in order to assess the effectiveness of DBCIs in small research studies. We exemplify the hypotheses and evaluation methods based on our experience from the CAncer PAtients Better Life Experience(\href{https://capable-project.eu/}{CAPABLE}) Project \cite{parimbelli2021cancer}.

\section{Related Work}

Yardley et al.  \cite{yardley2016understanding}  introduced the concept of micro- and macro-level engagement to understand user interaction, usage, and behavior changes related to interventions. Micro-level engagement involves immediate interaction with the intervention, such as usage frequency and user interest. In contrast, macro-level engagement focuses on the user's commitment to the behavior change process, like their motivation to alter behavior (where motivation moves from extrinsic to intrinsic), and is tied to the intervention's behavioral goals. Users may start with micro-level engagement, characterized by frequent interaction, but may shift towards macro-level engagement, persisting in their behavior change efforts even as direct interaction with the platform decreases. Thus, rather than advocating for sustained, high levels of interaction with digital interventions, the focus should be on understanding what constitutes \textit{'effective engagement'} namely, the level of engagement required to attain the desired outcomes \cite{michie2017developing}.

Short et al. \cite{short2018measuring} cataloged a comprehensive array of methodologies for gauging engagement, spanning qualitative metrics, self-reported questionnaires, ecological momentary assessments, analytics from system usage, sensor-derived data, social media analytics, and psychophysiological indicators. They advocated for a multifaceted approach to accurately capture user engagement at both the micro and macro levels. This entails the strategic selection of measurement tools, with questionnaires being more conducive for macro-level engagement analysis and system usage data offering insights into micro-level engagement.

In our methodology, we assess engagement on both levels: micro is evaluated using self-reported data from mobile apps and smartwatch sensor data for real-world insights, while macro is gauged through questionnaires (see Section \ref{engagment}). 

\section{The CAPABLE DBCI study}

The CAPABLE project developed a mobile health (mHealth) application which includes multiple DBCIs aiming to enhance the well-being of cancer patients. We examine the impact of engagement (see Section \ref{engagment}) with mobile BCIs (see Section \ref{BCI}) on the quality of life (QoL) and functioning (see Section \ref{OM}) of this population (see Section \ref{population}).

\subsection{Population}
\label{population}
\begin{figure}
    \centering
    \begin{subfigure}[b]{0.29\textwidth}
        \centering
        \includegraphics[width=\textwidth]{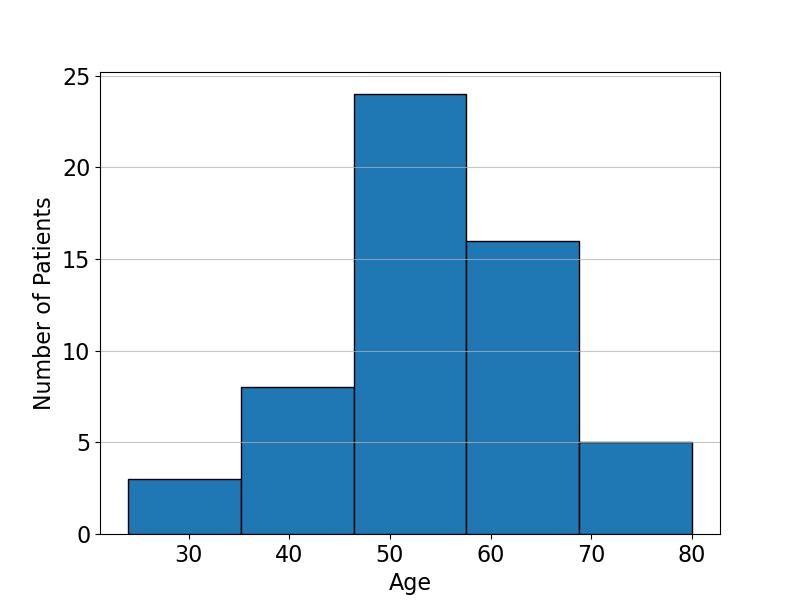}
        \caption{Age Distribution}
        \label{fig:age_dist}
    \end{subfigure}
    \begin{subfigure}[b]{0.29\textwidth}
        \centering
        \includegraphics[width=\textwidth]{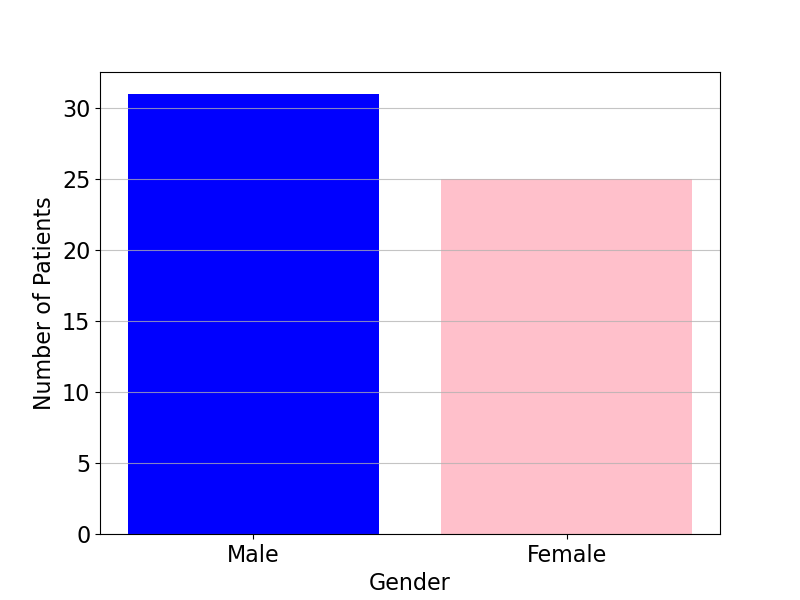}
        \caption{Gender Distribution}
        \label{fig:gender_dist}
    \end{subfigure}
 \begin{subfigure}[b]{0.29\textwidth}
        \centering
        \includegraphics[width=\textwidth]{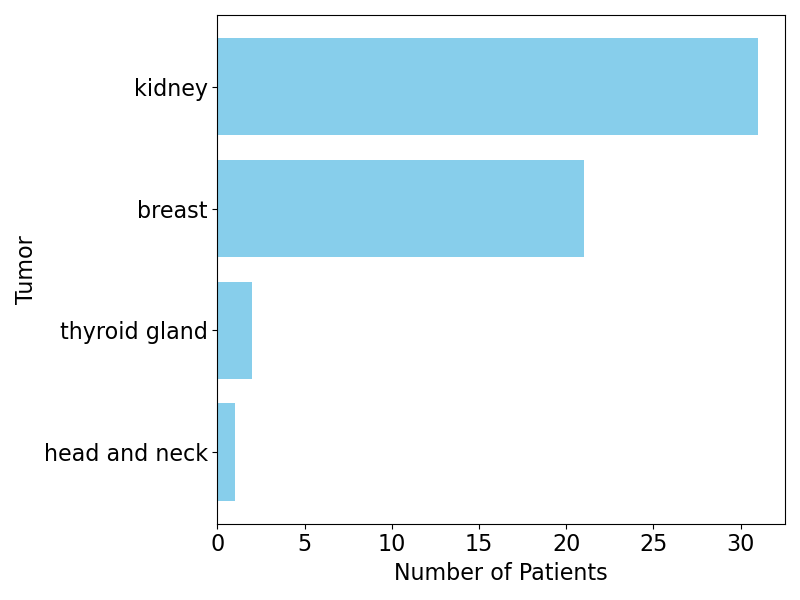}
        \caption{Tumor types}
        \label{fig:tumor}
          \end{subfigure}
     \caption{Patient Demographic and Cancer Distribution}
    \label{fig:age_gender_dist}
\end{figure}

\begin{figure}
    \centering
    \begin{subfigure}[b]{0.29\textwidth}
        \centering
        \includegraphics[width=\textwidth]{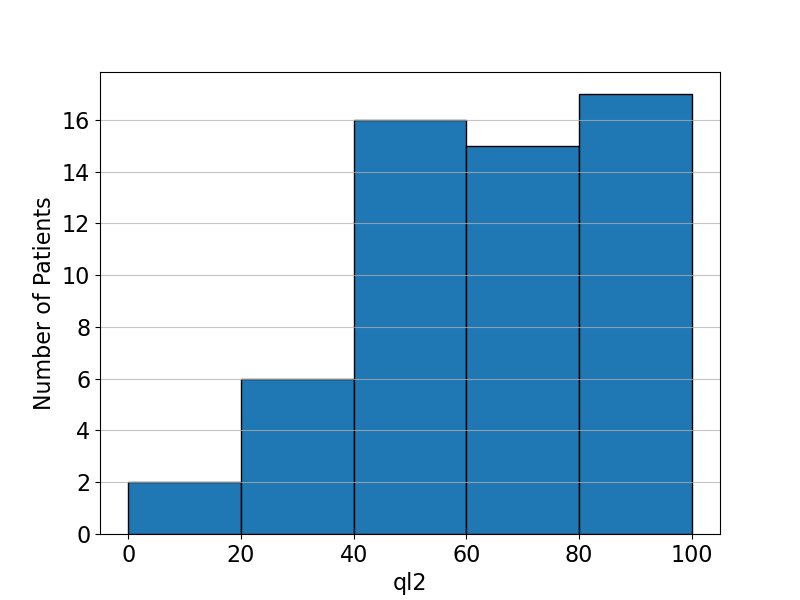}
        \caption{General Quality of Life}
        \label{fig:enrolment_qol}
    \end{subfigure}
    \begin{subfigure}[b]{0.29\textwidth}
        \centering
        \includegraphics[width=\textwidth]{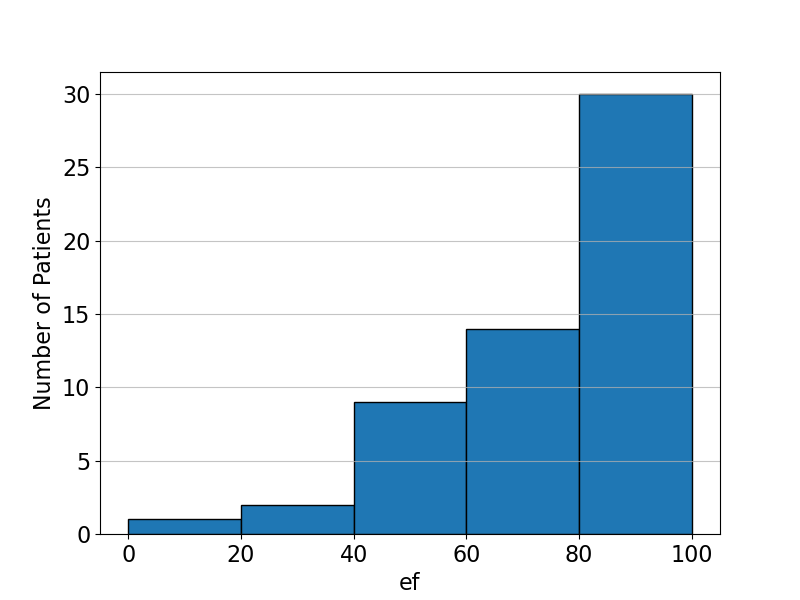}
        \caption{Emotional Functioning}
        \label{fig:enrolment_ef}
    \end{subfigure}
    \begin{subfigure}[b]{0.29\textwidth}
        \centering
        \includegraphics[width=\textwidth]{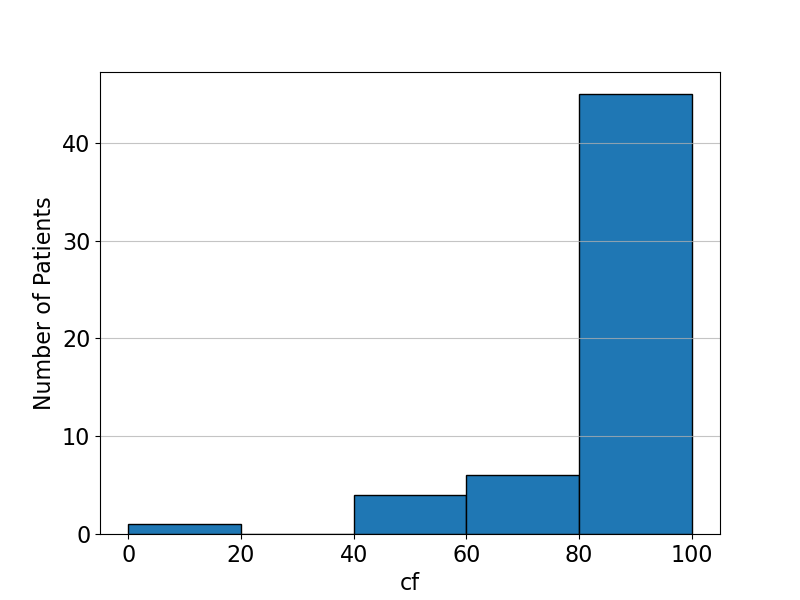}
        \caption{Cognitive Functioning}
        \label{fig:enrolment_cf}
    \end{subfigure}
    \begin{subfigure}[b]{0.29\textwidth}
        \centering
        \includegraphics[width=\textwidth]{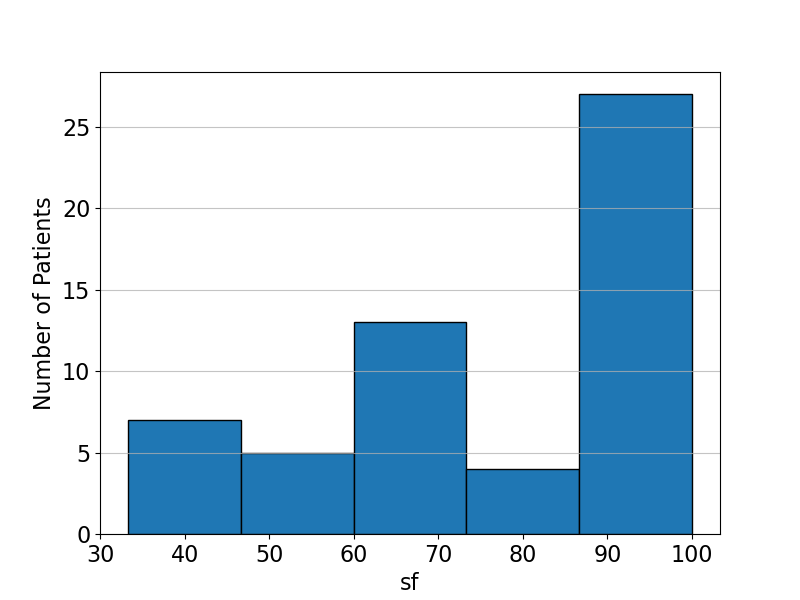}
        \caption{Social Functioning}
        \label{fig:enrolment_sf}
    \end{subfigure}
    \begin{subfigure}[b]{0.29\textwidth}
        \centering
        \includegraphics[width=\textwidth]{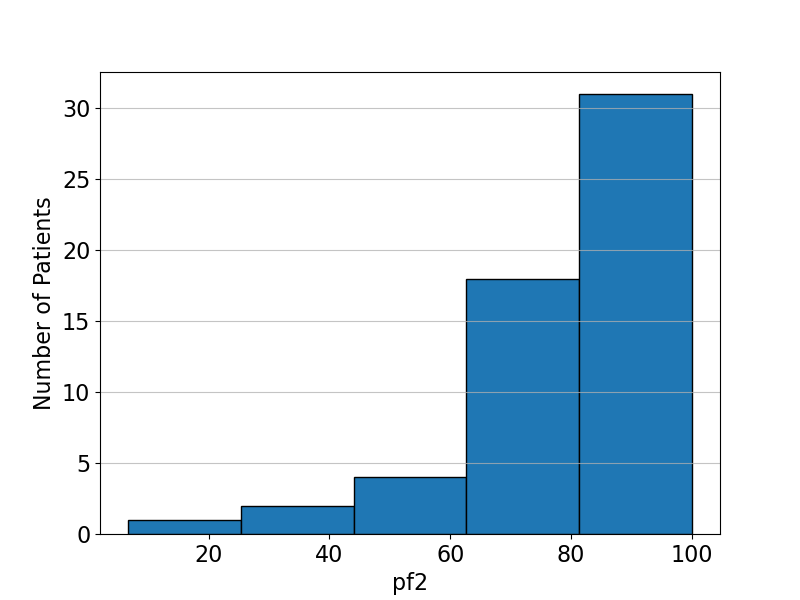}
        \caption{Physical Functioning}
        \label{fig:enrolment_pf}
    \end{subfigure}
    \begin{subfigure}[b]{0.29\textwidth}
        \centering
        \includegraphics[width=\textwidth]{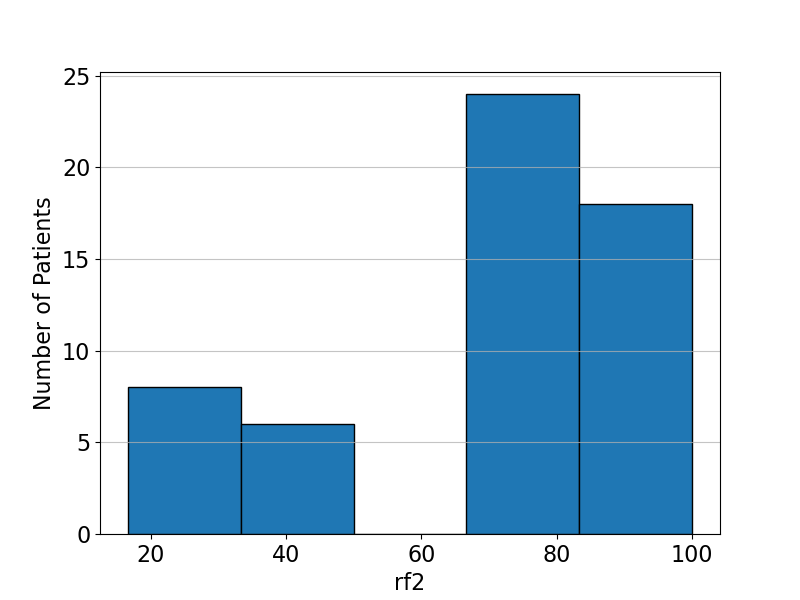}
        \caption{Role Functioning}
        \label{fig:enrolment_rf}
    \end{subfigure}
    \caption{Enrolment QoL scores}
    \label{fig:qol}
\end{figure}

Patients were enrolled in the CAPABLE study (ethical approval 2640 CE and NCT06161233 clinical trail) at the hospitals in Pavia (34 patients) and Bari (22 patients). The majority of these patients were in their 50s, as shown in Figure \ref{fig:age_dist}, and were most commonly diagnosed with malignant kidney or breast tumors, detailed in Figure \ref{fig:tumor}.  Notably, at enrollment, the patients' self reported  function and QoL tended to be skewed towards higher values, as depicted in Figure \ref{fig:qol}.

\subsection{Available DBCIs}
\label{BCI}
\begin{figure}[ht]
\centering
\digraph[scale=0.5]{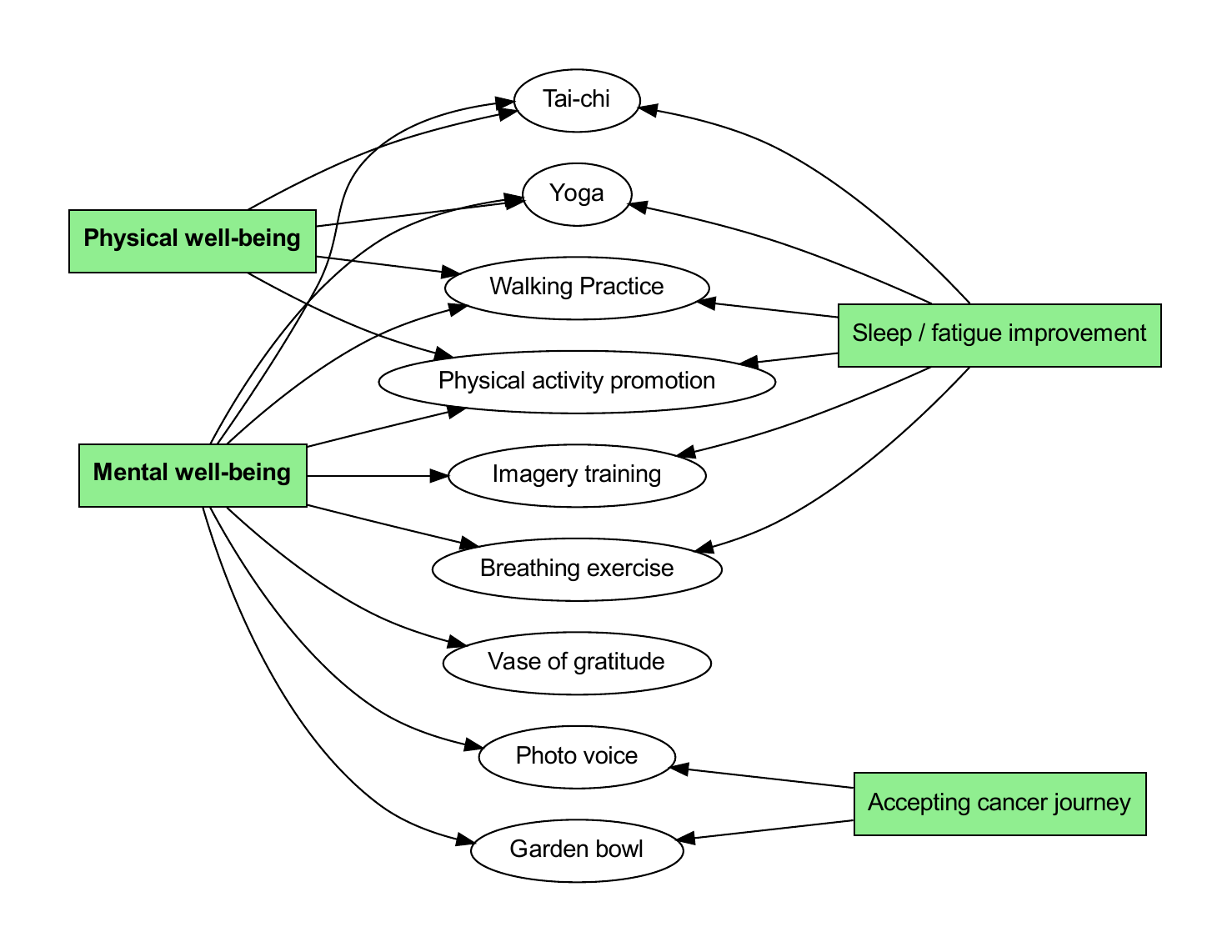}{
    rankdir=LR;
    node [fontname=arial];
    physical_wellbeing [label=<<b>Physical well-being</b>>, shape=rectangle, fillcolor=lightgreen, style=filled];
    mental_wellbeing [label=<<b>Mental well-being</b>>, shape=rectangle, fillcolor=lightgreen, style=filled];
    sleep_fatigue [label="Sleep / fatigue improvement", shape=rectangle, fillcolor=lightgreen, style=filled];
    cancer_journey [label="Accepting cancer journey", shape=rectangle, fillcolor=lightgreen, style=filled];
    
    tai_chi [label="Tai-chi"];
    yoga [label="Yoga"];
    walk [label="Walking Practice"];
    breathing [label="Breathing exercise"];
    imagery_training [label="Imagery training"];
    garden_bowl [label="Garden bowl"];
    photo_voice [label="Photo voice"];
    gratitude [label="Vase of gratitude"];
    pa_promo [label="Physical activity promotion"];
    
    physical_wellbeing -> {tai_chi, yoga, walk, pa_promo};
    mental_wellbeing -> {tai_chi, yoga, walk, breathing, imagery_training, garden_bowl, photo_voice, gratitude, pa_promo};
    sleep_fatigue -> {tai_chi, yoga, walk, breathing, imagery_training, pa_promo};
    cancer_journey -> {garden_bowl, photo_voice};
    {rank=source; physical_wellbeing; mental_wellbeing};
    {rank=sink; sleep_fatigue; cancer_journey};
}
\caption{Goals and capsules (general goals are located on the left, and specific goals are the right, PA = physical activity) }
\label{fig:goals_and_capsules}
\end{figure}

Following the European Society for Medical Oncology guidelines for managing cancer-related fatigue\cite{fabi2020cancer}, the CAPABLE app incorporates mindfulness-based stress reduction techniques, such as deep breathing, yoga, and walking in nature, alongside recommended physical exercises like Tai-chi, as supported by evidence-based meta-reviews\cite{muradnew}. Additionally, it offers horticultural therapy (garden bowl) and a photo-voice journal to help users express and accept their cancer journey. These activities, when performed regularly, are considered Behavior Change Interventions (BCIs) aimed at improving well-being. Through the app's augmentation with reminders, instructions, and additional features like talking pictures and commenting capabilities, they are transformed into Digital BCIs (DBCI). We refer to these DBCIs as \textit{virtual capsules} (or \textit{capsules} in short), drawing an analogy to pharmacological interventions where pills or medications are prescribed. The capsules can be  prescribed by clinicians or used by patients without specific recommendations and are linked with the patient's well-being goals \cite{lisowska2023sato}, such as mental well-being improvement (see Figure \ref{fig:goals_and_capsules}).

The prescription of capsules is associated with recommended weekly frequencies (times per week), which vary between capsules (see Table \ref{tab:capsule_frequencies}). 

\begin{table}[ht]
    \centering
    \small
    \begin{tabular}{lr}
        \toprule
        Capsule & Frequency \\
        \midrule
         Tai-chi & 7 \\
         Yoga & 7 \\
         Breathing exercise & 7 \\
         Physical activity promotion & 7 \\
         Walking Practice & 3 \\
         Imagery training & 1 \\
         Garden bowl & 1 \\
         Vase of gratitude & 1 \\
         Photo voice & 1 \\
         \bottomrule
    \end{tabular}
    \caption{Recommended weekly frequencies for specific capsules}
    \label{tab:capsule_frequencies}
\end{table}
\subsection{Engagement Measures}
\label{engagment}

In assessing user micro level engagement with DBCIs, our focus is primarily on self-reported feedback on capsule performance. For Walking Practice capsule we additionally use activity data automatically captured by a smartwatch to measure micro engagement with this intervention more objectively.
Given the intervention's emphasis on physical activity, we gauge macro-level engagement by assessing changes in attitude and motivation for exercise using the Behavioral Regulation in Exercise Questionnaire (BREQ-2). Utilizing responses from the BREQ-2, we calculated the Relative Autonomy Index (RAI), a composite score reflecting the extent of self-determined motivation.

Drawing on Michie et al. \cite{michie2017developing}, we define 'effective engagement' with capsules as a micro-level engagement that supports maintenance or improvement of initial well-being levels, as measured by functioning scores from the QLQ-C30 (see Section \ref{OM}).

\subsection{Outcome Measures}
\label{OM}
In our study, we focus on assessing quality of life using the EORTC QLQ-C30 questionnaire \cite{aaronson1993european}.  The European Organisation for Research and Treatment of Cancer (EORTC) Quality of Life Questionnaire (QLQ) C30 assesses two key dimensions of patient well-being: functional domains (including physical, emotional, social, role and cognitive functioning) and symptom domains. In the functional domains, higher scores indicate better functioning, whereas in the symptom domains, lower scores denote a more favorable outcome.

At baseline, participants were required to complete the QLQ-C30 and BREQ-2 questionnaires. Follow-up assessments using the QLQ-C30 were conducted at 3 and 6 months. The research particularly focuses on the trajectory of these metrics from enrollment to the 6-month endpoint, offering insights into the DBCIs impact.

\subsection{Data Collection and Storage}
\label{sec:data_collection}

The data was collected in Observational Medical Outcomes Partnership Common Data Model (OMOP-CDM) standard and made accessible via HL7 Fast Healthcare Interoperability Resources (FHIR) interfaces \cite{lanzola2023case}. The data was captured through four distinct avenues: (1) via a mobile patient app (including patient-reported outcomes such as Patient Health Questionnaire-9 (PHQ9) for assessing depression, Generalised Anxiety Disorder Assessment (GAD7), and self-reported duration of walks and symptoms), (2) through an ASUS smartwatch (capturing metrics like the number of steps or heart rate) (3) during clinical visits using the CAPABLE physician dashboard,
and (4) via manually completed enrollment questionnaires (QLQ-C30, BREQ2), the results of which were stored on the RedCap platform. The frequency of data collection varied: patient-reported questionnaires like QLQ-C30 were captured every three months, others like BREQ-2 at enrollment and at the end of the study, while activity and physiological metrics were captured automatically in minute-by-minute time steps.

\section{A Methodology for Assessing the Effectiveness of DBCIs for a Small Study Population}

The proposed methodology consists of a sequence of seven research questions (RQs). The first 4 questions are linked to micro-engagement and the last 3 to both micro- and macro-engagement.

\subsection{Capsule Prescription vs Autonomous Engagement}
\label{sec:capsule_prescription}
 
\textbf{RQ1: Do clinicians prescribe capsules, and how frequently do patients independently engage with capsule without doctor's explicit recommendations?}

\begin{figure}[ht]
    \centering
    \begin{subfigure}[b]{0.49\textwidth}
        \centering
        \includegraphics[width=\textwidth]{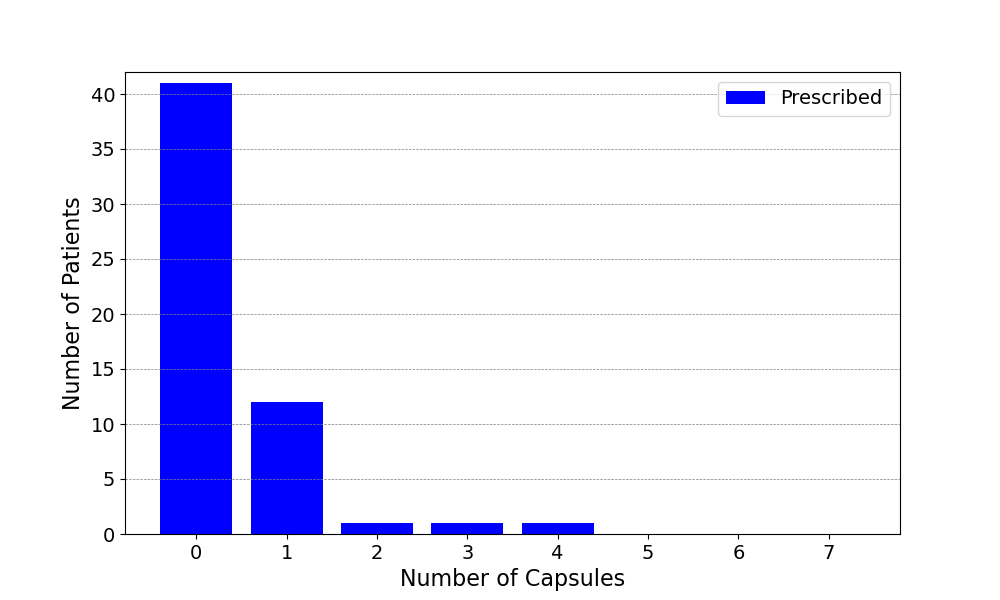}
        \caption{No. capsules clinician prescribed per patients} 
        \label{fig:num_prescriptions}
    \end{subfigure}
     \begin{subfigure}[b]{0.49\textwidth}
        \centering
        \includegraphics[width=\textwidth]{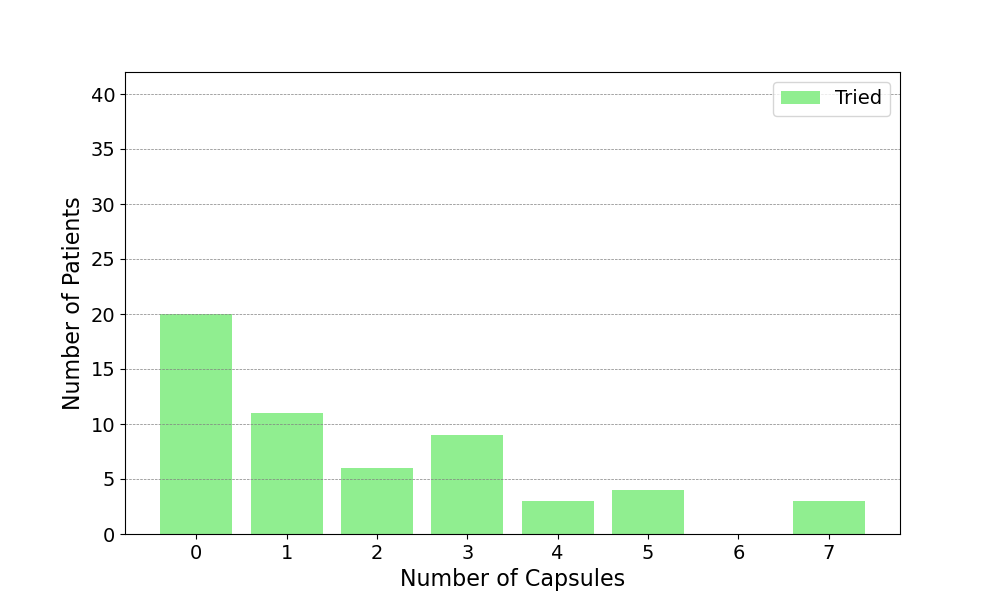}
          \centering
        \caption{No. capsules patients engaged with at least once} 
        \label{fig:num_patient_capsules}
    \end{subfigure}
    \caption{Capsules Statistic}
  
\end{figure}

We aim to uncover both the clinical practices in prescribing DBCIs and the autonomous actions of patients in trying such treatments. Data show a varied approach to prescribing: 13 patients were advised to walk for 30 minutes three times weekly, while  garden bowl, photo voice, and physical activity promotion (i.e., videos with physical exercises), Yoga and Tai-chi were prescribed to one patient each.  These observations are detailed in Table \ref{table:engagement_metrics}, which presents the number of patients who reported engaging with each intervention at least once. Although only 15 out of 56 patients were directly prescribed one or more interventions (see Figure \ref{fig:num_prescriptions}), a total of 36 patients reported at least one engagement with one of the interventions, i.e., tried a capsule. Many patients tried capsules on their own, with some trying up to 7 different capsules (see Figure \ref{fig:num_patient_capsules}). The number of capsules prescribed does not correlate with the number of capsules tried by patients (r=-0.08, $p>0.05$).

\begin{table}[ht]
\centering
\small 
\begin{tabular}{
>{\raggedright\arraybackslash}p{3.2cm}
>{\centering\arraybackslash}p{1.3cm}
>{\centering\arraybackslash}p{1.5cm}
>{\centering\arraybackslash}p{2.0cm}
>{\centering\arraybackslash}p{2.6cm}
}
\toprule
Capsule & Reported & Prescribed & Prescribed \& Reported & Not Prescribed \& Reported \\
\midrule
Vase of Gratitude & 15 & 0 & 0 & 15  \\
Garden Bowl  & 4 & 1 & 0 & 4 \\
Imagery Training & 15 & 0 & 0 & 15  \\
Yoga & 1 & 1 & 0 & 1 \\
Photo Voice  & 17 & 1 & 1 & 16  \\
PA Videos  & 7 & 1 & 0 & 7  \\
Breathing Exercise & 21 & 0 & 0 & 21  \\
Tai-chi & 2 & 1 & 0 & 2  \\
Walking Practice & 21 & 13 & 7 & 14  \\
\bottomrule
\end{tabular}
\caption{The table presents the number of patients who reported at least once engagement with the capsule (i.e., Tried) or had the capsule prescribed. It also highlights those who reported on the capsule without having it prescribed. Note that a prescription of a capsule does not automatically equate to patients reporting on its performance. Conversely, the absence of a prescription does not preclude patients from trying it out.}
\label{table:engagement_metrics}
\end{table}

\begin{figure}[ht!]
    \centering
    \begin{subfigure}{.48\textwidth}
        \centering
        \includegraphics[width=1.0\linewidth]{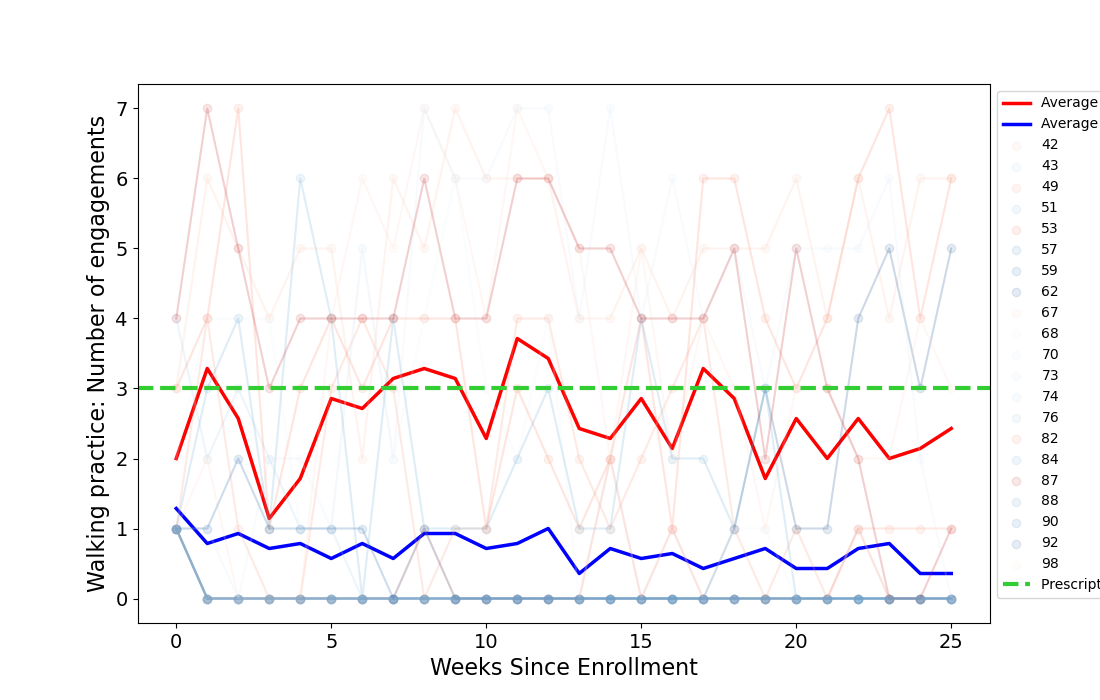}
        \caption{Weekly walks based on self-report}
        \label{walkreport}
    \end{subfigure}%
    \begin{subfigure}{.48\textwidth}
        \centering
        \includegraphics[width=1.0\linewidth]{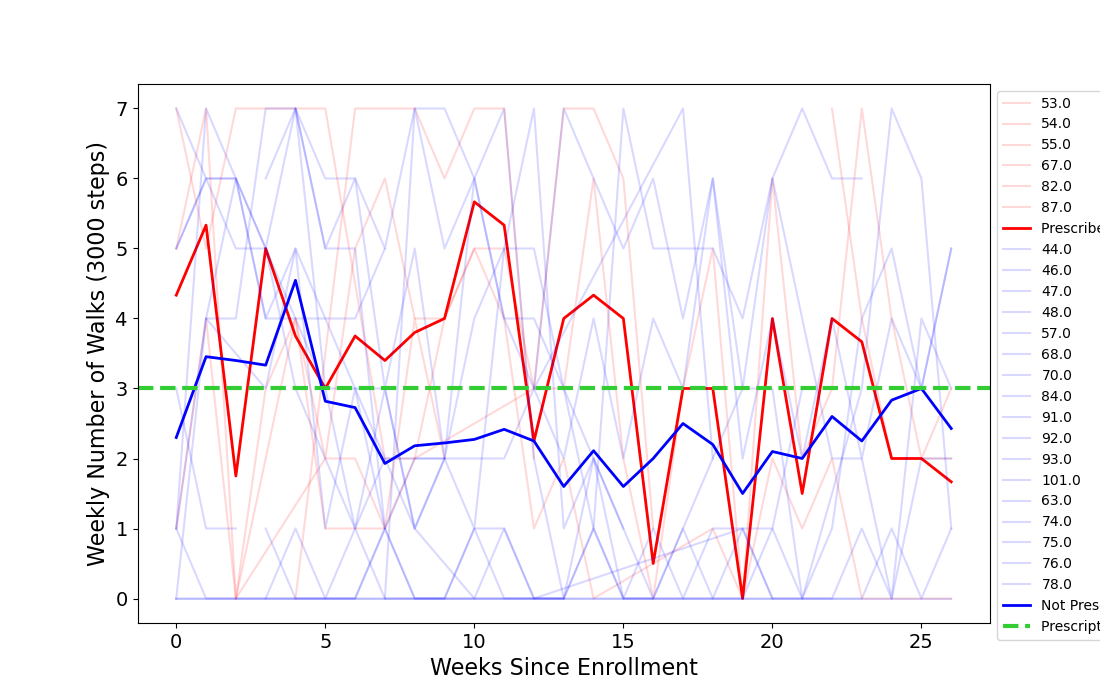}

        \caption{Weekly walks based of watch data}
        \label{walkwatch}
         \end{subfigure}
    \begin{subfigure}{1.0\textwidth}
        \centering
        \includegraphics[width=0.9\linewidth]{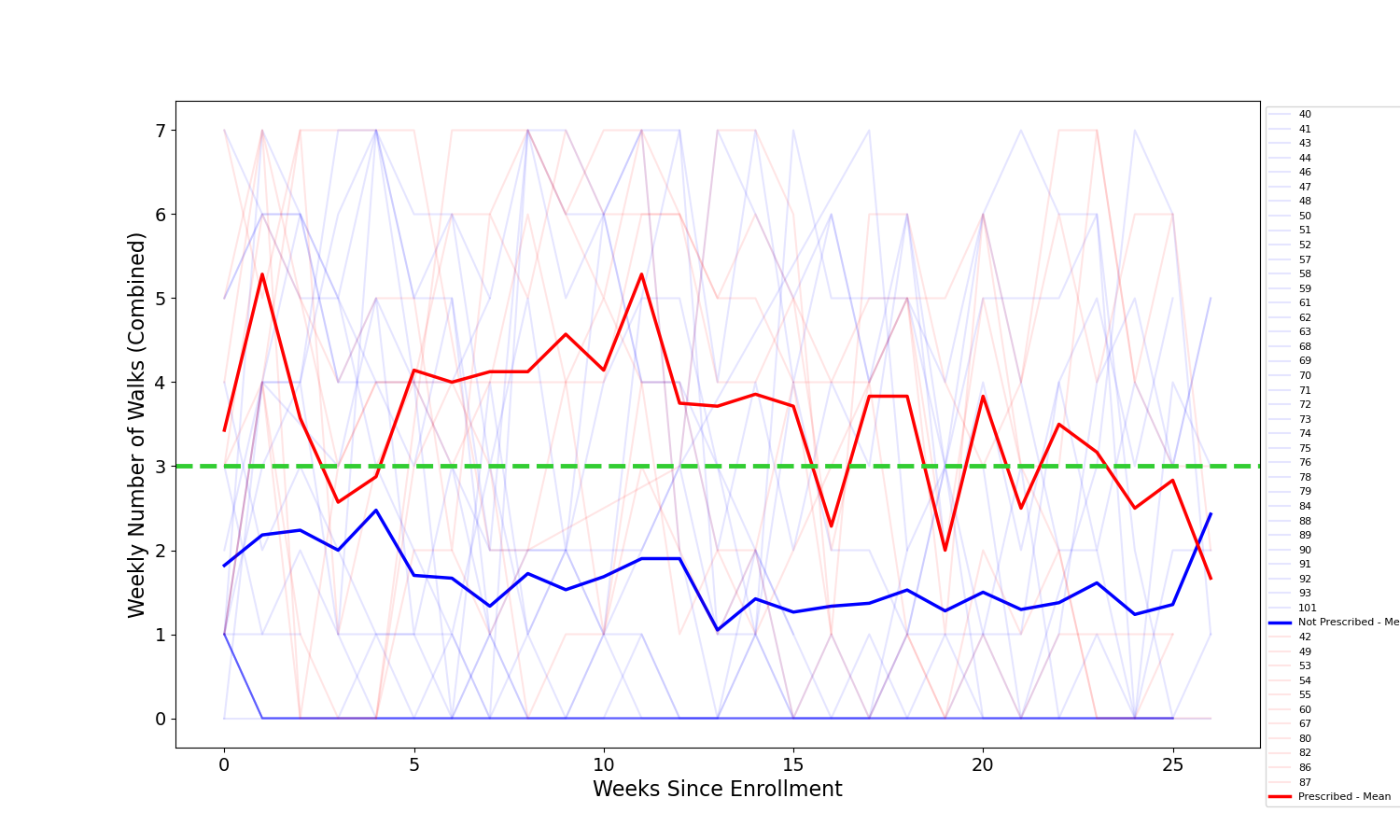}

        \caption{Weekly walks based on the union of self report and watch data. }
        \label{walkcombined}
         \end{subfigure}
    \caption{Engagement with Walking Capsule. Patients prescribed the capsule are represented in red, while those who engaged with the capsule without a specific prescription from clinicians are shown in blue. For those prescribed the capsule, the recommended weekly dosage is indicated by a green dashed line.  }
\end{figure}

\begin{table}
\centering
\resizebox{\textwidth}{!}{
\begin{tabular}{cllrrl}
\hline
\multicolumn{1}{|c|}{\textbf{RQ}} &
  \multicolumn{1}{l|}{\textbf{Dependent variable (DV)}} &
  \multicolumn{1}{l|}{\textbf{Patient Group}} &
  \multicolumn{1}{l|}{\textbf{No. Patients}} &
  \multicolumn{1}{l|}{\textbf{DV Mean}} &
  \multicolumn{1}{l|}{\textbf{p}} \\ \hline
\multicolumn{1}{|c|}{} &
  \multicolumn{1}{l|}{} &
  \multicolumn{1}{l|}{Prescribed} &
  \multicolumn{1}{r|}{15} &
  \multicolumn{1}{r|}{41.15} &
  \multicolumn{1}{c|}{} \\ \cline{3-5}
\multicolumn{1}{|c|}{} &
  \multicolumn{1}{l|}{} &
  \multicolumn{1}{l|}{Non-prescribed} &
  \multicolumn{1}{r|}{32} &
  \multicolumn{1}{r|}{12.97} &
  \multicolumn{1}{c|}{\multirow{-2}{*}{\textgreater{}0.05}} \\ \cline{3-6} 
\multicolumn{1}{|c|}{} &
  \multicolumn{1}{l|}{} &
  \multicolumn{1}{l|}{\cellcolor[HTML]{D9EAD3}Prescribed (Engaged )} &
  \multicolumn{1}{r|}{\cellcolor[HTML]{D9EAD3}7} &
  \multicolumn{1}{r|}{\cellcolor[HTML]{D9EAD3}76.43} &
  \multicolumn{1}{c|}{\cellcolor[HTML]{D9EAD3}} \\ \cline{3-5}
\multicolumn{1}{|c|}{} &
  \multicolumn{1}{l|}{\multirow{-4}{*}{\begin{tabular}[c]{@{}l@{}}Total Self-reported Engagements \\ Across All Capsules (6 months)\end{tabular}}} &
  \multicolumn{1}{l|}{\cellcolor[HTML]{D9EAD3}Non-prescribed (Engaged)} &
  \multicolumn{1}{r|}{\cellcolor[HTML]{D9EAD3}26} &
  \multicolumn{1}{r|}{\cellcolor[HTML]{D9EAD3}15.96} &
  \multicolumn{1}{c|}{\multirow{-2}{*}{\cellcolor[HTML]{D9EAD3}0.002}} \\ \cline{2-6} 
\multicolumn{1}{|c|}{} &
  \multicolumn{1}{l|}{} &
  \multicolumn{1}{l|}{Prescribed Walk} &
  \multicolumn{1}{r|}{11} &
  \multicolumn{1}{r|}{42.7} &
  \multicolumn{1}{c|}{} \\ \cline{3-5}
\multicolumn{1}{|c|}{} &
  \multicolumn{1}{l|}{} &
  \multicolumn{1}{l|}{Non-prescribed} &
  \multicolumn{1}{r|}{34} &
  \multicolumn{1}{r|}{7.35} &
  \multicolumn{1}{c|}{\multirow{-2}{*}{\textgreater{}0.05}} \\ \cline{3-6} 
\multicolumn{1}{|c|}{} &
  \multicolumn{1}{l|}{} &
  \multicolumn{1}{l|}{\cellcolor[HTML]{D9EAD3}Prescribed Walk (Engaged )} &
  \multicolumn{1}{r|}{\cellcolor[HTML]{D9EAD3}6} &
  \multicolumn{1}{r|}{\cellcolor[HTML]{D9EAD3}77.5} &
  \multicolumn{1}{c|}{\cellcolor[HTML]{D9EAD3}} \\ \cline{3-5}
\multicolumn{1}{|c|}{} &
  \multicolumn{1}{l|}{\multirow{-4}{*}{\begin{tabular}[c]{@{}l@{}}Total Self-reported Engagements\\  Walk Practice (6 months)\end{tabular}}} &
  \multicolumn{1}{l|}{\cellcolor[HTML]{D9EAD3}Non-prescribed (Engaged)} &
  \multicolumn{1}{r|}{\cellcolor[HTML]{D9EAD3}14} &
  \multicolumn{1}{r|}{\cellcolor[HTML]{D9EAD3}17.86} &
  \multicolumn{1}{c|}{\multirow{-2}{*}{\cellcolor[HTML]{D9EAD3}0.013}} \\ \cline{2-6} 
\multicolumn{1}{|c|}{} &
  \multicolumn{1}{l|}{} &
  \multicolumn{1}{l|}{Prescribed Walk \& Wore Watch} &
  \multicolumn{1}{r|}{6} &
  \multicolumn{1}{r|}{49.67} &
  \multicolumn{1}{l|}{} \\ \cline{3-5}
\multicolumn{1}{|c|}{} &
  \multicolumn{1}{l|}{\multirow{-2}{*}{Total Walks Smartwatch (6 months)}} &
  \multicolumn{1}{l|}{Non-prescribed \& Wore Watch} &
  \multicolumn{1}{r|}{16} &
  \multicolumn{1}{r|}{39.81} &
  \multicolumn{1}{l|}{\multirow{-2}{*}{\textgreater{}0.05}} \\ \cline{2-6} 
\multicolumn{1}{|c|}{} &
  \multicolumn{1}{l|}{} &
  \multicolumn{1}{l|}{\cellcolor[HTML]{D9EAD3}Prescribed Walk (Self-reported and/or Wore Watch)} &
  \multicolumn{1}{r|}{\cellcolor[HTML]{D9EAD3}8} &
  \multicolumn{1}{r|}{\cellcolor[HTML]{D9EAD3}80.75} &
  \multicolumn{1}{c|}{\cellcolor[HTML]{D9EAD3}} \\ \cline{3-5}
\multicolumn{1}{|c|}{\multirow{-12}{*}{RQ2}} &
  \multicolumn{1}{l|}{\multirow{-2}{*}{\begin{tabular}[c]{@{}l@{}}Total Walks (6 months)\\  Max(Self-report, Smartwatch)\end{tabular}}} &
  \multicolumn{1}{l|}{\cellcolor[HTML]{D9EAD3}Non-prescribed (Self-reported and/or Wore Watch)} &
  \multicolumn{1}{r|}{\cellcolor[HTML]{D9EAD3}23} &
  \multicolumn{1}{r|}{\cellcolor[HTML]{D9EAD3}35.78} &
  \multicolumn{1}{c|}{\multirow{-2}{*}{\cellcolor[HTML]{D9EAD3}0.009}} \\ \hline
\multicolumn{1}{|c|}{} &
  \multicolumn{1}{l|}{} &
  \multicolumn{1}{l|}{\cellcolor[HTML]{EFEFEF}Improved/Maintained ql2} &
  \multicolumn{1}{r|}{\cellcolor[HTML]{EFEFEF}17} &
  \multicolumn{1}{r|}{\cellcolor[HTML]{EFEFEF}2.68} &
  \multicolumn{1}{l|}{} \\ \cline{3-5}
\multicolumn{1}{|c|}{} &
  \multicolumn{1}{l|}{} &
  \multicolumn{1}{l|}{Decline ql2} &
  \multicolumn{1}{r|}{5} &
  \multicolumn{1}{r|}{3.31} &
  \multicolumn{1}{l|}{} \\ \cline{3-5}
\multicolumn{1}{|c|}{} &
  \multicolumn{1}{l|}{} &
  \multicolumn{1}{l|}{\cellcolor[HTML]{EFEFEF}Improved/Maintained ef} &
  \multicolumn{1}{r|}{\cellcolor[HTML]{EFEFEF}12} &
  \multicolumn{1}{r|}{\cellcolor[HTML]{EFEFEF}2.87} &
  \multicolumn{1}{l|}{} \\ \cline{3-5}
\multicolumn{1}{|c|}{} &
  \multicolumn{1}{l|}{} &
  \multicolumn{1}{l|}{Decline ef} &
  \multicolumn{1}{r|}{10} &
  \multicolumn{1}{r|}{2.77} &
  \multicolumn{1}{l|}{} \\ \cline{3-5}
\multicolumn{1}{|c|}{} &
  \multicolumn{1}{l|}{} &
  \multicolumn{1}{l|}{\cellcolor[HTML]{EFEFEF}Improved/Maintained ph2} &
  \multicolumn{1}{r|}{\cellcolor[HTML]{EFEFEF}15} &
  \multicolumn{1}{r|}{\cellcolor[HTML]{EFEFEF}2.8} &
  \multicolumn{1}{l|}{} \\ \cline{3-5}
\multicolumn{1}{|c|}{} &
  \multicolumn{1}{l|}{} &
  \multicolumn{1}{l|}{Decline ph2} &
  \multicolumn{1}{r|}{7} &
  \multicolumn{1}{r|}{2.88} &
  \multicolumn{1}{l|}{} \\ \cline{3-5}
\multicolumn{1}{|c|}{} &
  \multicolumn{1}{l|}{} &
  \multicolumn{1}{l|}{\cellcolor[HTML]{EFEFEF}Improved/Maintained cf} &
  \multicolumn{1}{r|}{\cellcolor[HTML]{EFEFEF}20} &
  \multicolumn{1}{r|}{\cellcolor[HTML]{EFEFEF}2.87} &
  \multicolumn{1}{l|}{} \\ \cline{3-5}
\multicolumn{1}{|c|}{} &
  \multicolumn{1}{l|}{} &
  \multicolumn{1}{l|}{Decline cf} &
  \multicolumn{1}{r|}{2} &
  \multicolumn{1}{r|}{2.41} &
  \multicolumn{1}{l|}{} \\ \cline{3-5}
\multicolumn{1}{|c|}{} &
  \multicolumn{1}{l|}{} &
  \multicolumn{1}{l|}{\cellcolor[HTML]{EFEFEF}Improved/Maintained sf} &
  \multicolumn{1}{r|}{\cellcolor[HTML]{EFEFEF}18} &
  \multicolumn{1}{r|}{\cellcolor[HTML]{EFEFEF}2.72} &
  \multicolumn{1}{l|}{} \\ \cline{3-5}
\multicolumn{1}{|c|}{} &
  \multicolumn{1}{l|}{} &
  \multicolumn{1}{l|}{Decline sf} &
  \multicolumn{1}{r|}{4} &
  \multicolumn{1}{r|}{3.2} &
  \multicolumn{1}{l|}{} \\ \cline{3-5}
\multicolumn{1}{|c|}{} &
  \multicolumn{1}{l|}{} &
  \multicolumn{1}{l|}{\cellcolor[HTML]{EFEFEF}Improved/Maintained rf2} &
  \multicolumn{1}{r|}{\cellcolor[HTML]{EFEFEF}18} &
  \multicolumn{1}{r|}{\cellcolor[HTML]{EFEFEF}3} &
  \multicolumn{1}{l|}{} \\ \cline{3-5}
\multicolumn{1}{|c|}{} &
  \multicolumn{1}{l|}{\multirow{-12}{*}{Average Weekly Walks Smartwatch}} &
  \multicolumn{1}{l|}{Decline rf2} &
  \multicolumn{1}{r|}{4} &
  \multicolumn{1}{r|}{2.05} &
  \multicolumn{1}{l|}{} \\ \cline{2-5}
\multicolumn{1}{|c|}{} &
  \multicolumn{1}{l|}{} &
  \multicolumn{1}{l|}{$\geq$ 3 weekly walks (Watch)} &
  \multicolumn{1}{r|}{14} &
  \multicolumn{1}{r|}{0.44} &
  \multicolumn{1}{l|}{} \\ \cline{3-5}
\multicolumn{1}{|c|}{\multirow{-14}{*}{RQ3}} &
  \multicolumn{1}{l|}{\multirow{-2}{*}{Functioning Improvement}} &
  \multicolumn{1}{l|}{\textless{}3 weekly walks (Watch)} &
  \multicolumn{1}{r|}{8} &
  \multicolumn{1}{r|}{0.73} &
  \multicolumn{1}{l|}{\multirow{-14}{*}{\textgreater{}0.05}} \\ \hline
\multicolumn{1}{|c|}{} &
  \multicolumn{1}{l|}{} &
  \multicolumn{1}{l|}{RAI Improved} &
  \multicolumn{1}{r|}{9} &
  \multicolumn{1}{r|}{50.9} &
  \multicolumn{1}{r|}{} \\ \cline{3-5}
\multicolumn{1}{|c|}{\multirow{-2}{*}{RQ5}} &
  \multicolumn{1}{l|}{\multirow{-2}{*}{\begin{tabular}[c]{@{}l@{}}Total Self-reported Engagements \\ Across All Capsules (6 months)\end{tabular}}} &
  \multicolumn{1}{l|}{RAI Not improved} &
  \multicolumn{1}{r|}{12} &
  \multicolumn{1}{r|}{12.3} &
  \multicolumn{1}{r|}{\multirow{-2}{*}{0.068}} \\ \hline
\multicolumn{1}{l}{} &
  \textbf{Predictor} &
   &
  \multicolumn{1}{l}{} &
  \multicolumn{1}{l}{\textbf{Mean}} &
   \\ \hline
\multicolumn{1}{|c|}{} &
  \multicolumn{1}{l|}{} &
  \multicolumn{1}{l|}{\cellcolor[HTML]{EFEFEF}Improved} &
  \multicolumn{1}{r|}{\cellcolor[HTML]{EFEFEF}26} &
  \multicolumn{1}{r|}{\cellcolor[HTML]{EFEFEF}58} &
  \multicolumn{1}{r|}{\cellcolor[HTML]{D9EAD3}0.002} \\ \cline{3-6} 
\multicolumn{1}{|c|}{} &
  \multicolumn{1}{l|}{\multirow{-2}{*}{ql2 at enrollment}} &
  \multicolumn{1}{l|}{\cellcolor[HTML]{D9EAD3}Not improved} &
  \multicolumn{1}{r|}{\cellcolor[HTML]{D9EAD3}19} &
  \multicolumn{1}{r|}{\cellcolor[HTML]{D9EAD3}74} &
  \multicolumn{1}{r|}{\cellcolor[HTML]{D9EAD3}} \\ \cline{2-6} 
\multicolumn{1}{|c|}{} &
  \multicolumn{1}{l|}{} &
  \multicolumn{1}{l|}{\cellcolor[HTML]{EFEFEF}Improved} &
  \multicolumn{1}{r|}{\cellcolor[HTML]{EFEFEF}26} &
  \multicolumn{1}{r|}{\cellcolor[HTML]{EFEFEF}73} &
  \multicolumn{1}{r|}{\cellcolor[HTML]{D9EAD3}} \\ \cline{3-5}
\multicolumn{1}{|c|}{} &
  \multicolumn{1}{l|}{\multirow{-2}{*}{ef at enrollment}} &
  \multicolumn{1}{l|}{\cellcolor[HTML]{D9EAD3}Not improved} &
  \multicolumn{1}{r|}{\cellcolor[HTML]{D9EAD3}19} &
  \multicolumn{1}{r|}{\cellcolor[HTML]{D9EAD3}85} &
  \multicolumn{1}{r|}{\multirow{-2}{*}{\cellcolor[HTML]{D9EAD3}0.023}} \\ \cline{2-6} 
\multicolumn{1}{|c|}{} &
  \multicolumn{1}{l|}{} &
  \multicolumn{1}{l|}{\cellcolor[HTML]{EFEFEF}Improved} &
  \multicolumn{1}{r|}{\cellcolor[HTML]{EFEFEF}26} &
  \multicolumn{1}{r|}{\cellcolor[HTML]{EFEFEF}81} &
  \multicolumn{1}{r|}{\cellcolor[HTML]{D9EAD3}} \\ \cline{3-5}
\multicolumn{1}{|c|}{} &
  \multicolumn{1}{l|}{\multirow{-2}{*}{cf at enrollment}} &
  \multicolumn{1}{l|}{\cellcolor[HTML]{D9EAD3}Not improved} &
  \multicolumn{1}{r|}{\cellcolor[HTML]{D9EAD3}19} &
  \multicolumn{1}{r|}{\cellcolor[HTML]{D9EAD3}94} &
  \multicolumn{1}{r|}{\multirow{-2}{*}{\cellcolor[HTML]{D9EAD3}0.004}} \\ \cline{2-6} 
\multicolumn{1}{|c|}{} &
  \multicolumn{1}{l|}{} &
  \multicolumn{1}{l|}{\cellcolor[HTML]{EFEFEF}Improved} &
  \multicolumn{1}{r|}{\cellcolor[HTML]{EFEFEF}26} &
  \multicolumn{1}{r|}{\cellcolor[HTML]{EFEFEF}77} &
  \multicolumn{1}{l|}{} \\ \cline{3-5}
\multicolumn{1}{|c|}{} &
  \multicolumn{1}{l|}{\multirow{-2}{*}{pf2 at enrollment}} &
  \multicolumn{1}{l|}{Not improved} &
  \multicolumn{1}{r|}{19} &
  \multicolumn{1}{r|}{86} &
  \multicolumn{1}{l|}{\multirow{-2}{*}{\textgreater{}0.05}} \\ \cline{2-6} 
\multicolumn{1}{|c|}{} &
  \multicolumn{1}{l|}{} &
  \multicolumn{1}{l|}{\cellcolor[HTML]{EFEFEF}Improved} &
  \multicolumn{1}{r|}{\cellcolor[HTML]{EFEFEF}26} &
  \multicolumn{1}{r|}{\cellcolor[HTML]{EFEFEF}72} &
  \multicolumn{1}{r|}{\cellcolor[HTML]{D9EAD3}} \\ \cline{3-5}
\multicolumn{1}{|c|}{} &
  \multicolumn{1}{l|}{\multirow{-2}{*}{sf at enrollment}} &
  \multicolumn{1}{l|}{\cellcolor[HTML]{D9EAD3}Not improved} &
  \multicolumn{1}{r|}{\cellcolor[HTML]{D9EAD3}19} &
  \multicolumn{1}{r|}{\cellcolor[HTML]{D9EAD3}88} &
  \multicolumn{1}{r|}{\multirow{-2}{*}{\cellcolor[HTML]{D9EAD3}0.02}} \\ \cline{2-6} 
\multicolumn{1}{|c|}{} &
  \multicolumn{1}{l|}{} &
  \multicolumn{1}{l|}{\cellcolor[HTML]{EFEFEF}Improved} &
  \multicolumn{1}{r|}{\cellcolor[HTML]{EFEFEF}26} &
  \multicolumn{1}{r|}{\cellcolor[HTML]{EFEFEF}64} &
  \multicolumn{1}{r|}{\cellcolor[HTML]{D9EAD3}} \\ \cline{3-5}
\multicolumn{1}{|c|}{} &
  \multicolumn{1}{l|}{\multirow{-2}{*}{rf2 at enrollment}} &
  \multicolumn{1}{l|}{\cellcolor[HTML]{D9EAD3}Not improved} &
  \multicolumn{1}{r|}{\cellcolor[HTML]{D9EAD3}19} &
  \multicolumn{1}{r|}{\cellcolor[HTML]{D9EAD3}85} &
  \multicolumn{1}{r|}{\multirow{-2}{*}{\cellcolor[HTML]{D9EAD3}0.002}} \\ \cline{2-6} 
\multicolumn{1}{|c|}{} &
  \multicolumn{1}{l|}{} &
  \multicolumn{1}{l|}{\cellcolor[HTML]{EFEFEF}Improved (filled BREQ2)} &
  \multicolumn{1}{r|}{\cellcolor[HTML]{EFEFEF}17} &
  \multicolumn{1}{r|}{\cellcolor[HTML]{EFEFEF}30} &
  \multicolumn{1}{l|}{} \\ \cline{3-5}
\multicolumn{1}{|c|}{\multirow{-14}{*}{RQ7}} &
  \multicolumn{1}{l|}{\multirow{-2}{*}{RAI at enrollment}} &
  \multicolumn{1}{l|}{Not improved (filled BREQ2)} &
  \multicolumn{1}{r|}{8} &
  \multicolumn{1}{r|}{20} &
  \multicolumn{1}{l|}{\multirow{-2}{*}{\textgreater{}0.05}} \\ \hline
  
\end{tabular}}

\caption{Results from the Mann-Whitney U test, which checks for differences between groups, are highlighted in green to indicate statistically significant outcomes at a level with less than a 5\% chance of occurring by random variation.}
\label{tab:stat-diff}

\end{table}
\subsection{Engagement with Capsules: the Impact of Prescription}

\textbf{RQ2: Do patients engage more when the capsules were prescribed to them?}

Our analysis started by comparing engagement frequencies across capsules between patients prescribed at least one capsule and those with none, as shown in Table~\ref{table:engagement_metrics}. For all patients completing a 6-month follow-up, no significant engagement difference was found (\(p > 0.05\)). Yet, among 33 patients who tried at least one capsule, a notable difference in engagement emerged between prescribed and non-prescribed patients (\(p = 0.002\)) (Table~\ref{tab:stat-diff}).

Focusing on the walking capsule due to its frequent prescription (Table~\ref{table:engagement_metrics}),
we found no significant overall engagement difference between prescribed and non-prescribed patients (p=0.1) (Table~\ref{tab:stat-diff}). However, 'engaged' patients—those who attempted walking at least once—revealed a substantial discrepancy (p=0.013), with prescribed individuals averaging 78 walks over 26 weeks compared to 18 by non-prescribed. This variation suggests potential differences in self-reported engagement, prompting us to examine actual walking behavior via smartwatch data (Figure \ref{walkreport}).

Smartwatch data confirmed that prescribed patients typically achieved the advised three weekly 30-minute walks of over 3000 steps, underscoring adherence to recommendations (Figure \ref{walkwatch}).


Among 9 patients reporting and wearing watches, significant walking frequency correlations emerged for three: two positive (r = 0.64, p = 0.001 and r = 0.9, p = 0.04) and one negative (r = -0.67, p = 0.003), suggesting discrepancies in self-reporting versus device data. This discrepancy implies some might report walks only when not wearing the device and omit them otherwise. Combining walk counts from both sources, assuming zeros for missing reports, Figure \ref{walkcombined} shows a significant difference in total walks between prescribed and non-prescribed groups (p=0.009), indicating prescription status influences engagement.



\subsection{Association Between Capsule Engagement and Quality of Life Outcomes}

\textbf{RQ3: Is the engagement with capsules linked with improved or sustained QoL?}

\begin{figure}[h!]
    \centering
    \begin{subfigure}{.8\textwidth}
        \centering
        \includegraphics[width=1.0\linewidth]{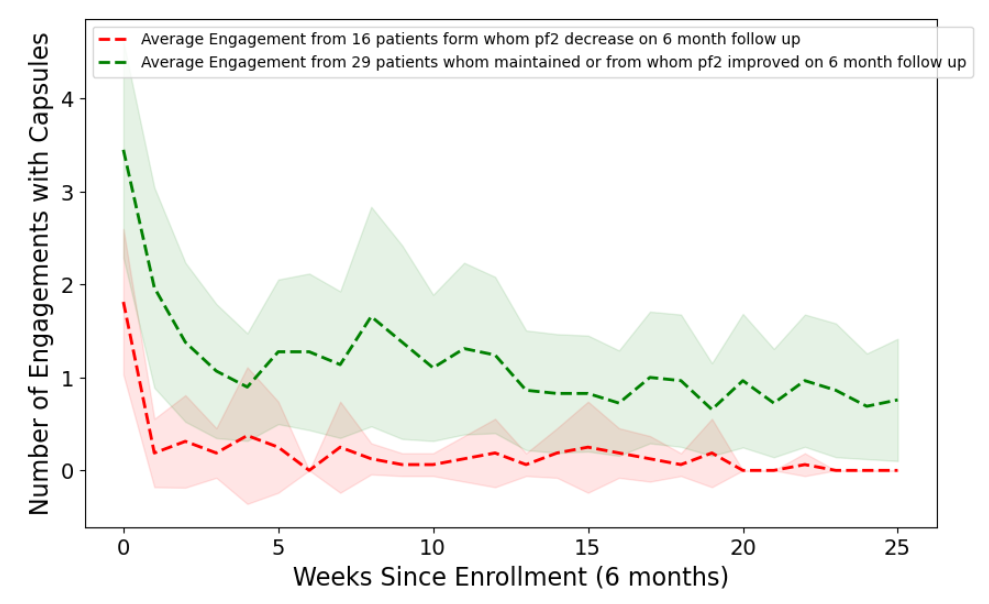}
        \caption{physical functioning (ph2)}
        \label{ph}
    \end{subfigure}%
    \newline
    \begin{subfigure}{.8\textwidth}
        \centering
        \includegraphics[width=1.0\linewidth]{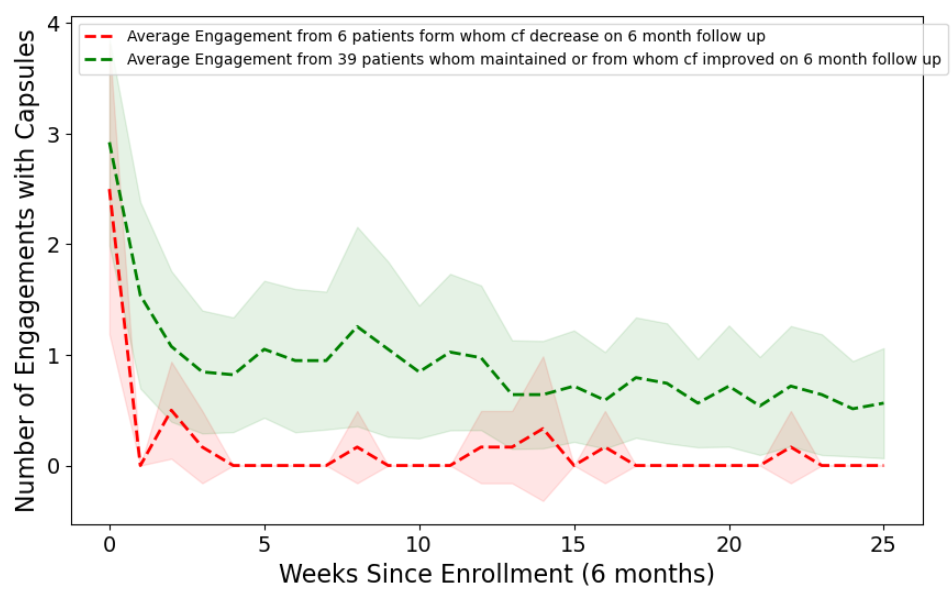}
        \caption{cognitive functioning (cf)}
        \label{cf}
         \end{subfigure}
    \caption{Engagement across all capsules for patients who maintained functioning (green) and those who declined (red)}
    \label{engagmentandotucome}
\end{figure}

This analysis explores the link between capsule engagement and quality of life (QoL) improvements or maintenance. Through self-reported data, we assess weekly engagement with capsules. Figure \ref{engagmentandotucome} shows patients maintaining or improving their QoL engaged with at least one capsule weekly. Consistent engagement correlated with sustained or enhanced physical and cognitive functioning. Conversely, those with declining functional status showed minimal engagement.


\textbf{RQ4: Is there evidence to establish causality between engagement with capsules and improvements in QoL?}

This inquiry aimed to discern whether a causal relationship exists between engaging in behaviour change interventions, specifically walking, and enhancements in QoL metrics. Given the limitations of self-reported data, the analysis sought to utilize data from wearable devices for a more objective measure of engagement. A challenge encountered was the initially high levels of functioning among many participants, which inherently limited the potential for observable improvement.

We focused on a subset of patients who consistently wore smartwatches throughout the 26-week intervention period. This subset comprised 22 patients, with 14 meeting the criterion of walking at least three times weekly and 8 not meeting this criterion. The functional improvement metric was calculated as follows:

\begin{equation}
\Delta F^i = \frac{F_{\text{Score6M}}^i}{F_{\text{ScoreEnrollment}}^i} - 1
\end{equation}

where \(i \in \{rl2, ql2, ef, pf2, cf, sf\}\), representing different functioning measures from the QLQ-C30, including quality of life (\(ql2\)), emotional (\(ef\)), physical (\(pf2\)), cognitive (\(cf\)), social (\(sf\)), and role (\(rl2\)) functioning.

We found no significant differences in functional improvement scores between the two groups of patients ($p>0.05$). Consequently, it was not possible to establish causality between engagement in walking and improvements in QoL. Similarly, the data did not support the hypothesis that higher initial QoL scores would lead to greater engagement with the prescribed walking regimen.

\subsection{Micro- vs Macro-level engagement}

\textbf{RQ5: Is micro- level engagement linked with changes in macro-level engagement?}

We explored the relationship between micro-level engagement (interaction with specific capsules) and macro-level engagement changes (overall changes in motivation). We assessed RAI score differences from the first to the last questionnaire, completed by 21 patients over a minimum five-month interval. Of these, 4 were prescribed the walking capsule, and 1 had contraindications for yoga and Tai-chi. Spearman's analysis showed a significant, positive correlation between RAI change and total capsule engagements (r=0.49, p=0.02), indicating micro-level engagement correlates with shifts in motivation. Specifically, engagement with the walking capsule alone also showed a significant positive correlation (r=0.46, p=0.03).

Patients were then categorized based on whether their RAI increased or not. Analyzing total capsule engagements, we observed that patients with increased self-determination engaged on average 50 times in 6 months, roughly twice per week. This contrasts with the average engagement of once a week for maintaining physical function, as seen in Figure \ref{ph}. Thus, transitioning from extrinsic to intrinsic motivation for capsule activities may necessitate more frequent engagement, though differences in total engagements between groups were not statistically significant (See Table \ref{tab:stat-diff})

\textbf{RQ6: Is the initial RAI linked with outcome improvement or changes in macro-level engagement?} 

We define the overall improvement in functioning as the total change across various domains, calculated using the following formula:
\begin{equation}
\text{TotalChange} = \sum_{i \in \{rl2, ql2, ef, pf2, cf, sf\}} \Delta F^i
\label{eq:tc}
\end{equation}

Twenty-seven patients completed the BREQ-2 questionnaire at least once. Our analysis found no correlation between the initial RAI and any of the functional outcomes, with all p-values greater than 0.05. Similarly, there was no significant relationship between the initial RAI and functional improvement ($p>0.05$), nor was the change in RAI significantly associated with total funtioning improvement as defined in equation \ref{eq:tc}.

\subsection{Predicting Intervention Outcome}

\textbf{RQ7: Is it possible to predict total improvement in functioning based on enrollment questioners, number of capsules prescribed, and demographic data?}

To distinguish between patients who overall improved or declined, we apply thresholding to the total sum to calculate the ``ImprovementIndicator'':
\begin{equation}
\text{ImprovementIndicator} = 
\begin{cases} 
1 & \text{if } \text{TotalChange} > 0 \\
0 & \text{otherwise}
\end{cases}
\end{equation}
Among the 45 patients for whom we had data at the 6-month follow-up, 26 were categorized as improved, and 19 were not improved (this includes 2 whose total functioning did not change and 17 who declined).

First, we utilized the AutoML AutoGluon package\cite{agtabular} to automatically identify a model that could perform well on this task. Based on AutoGluon models performance leaderboard we selected the Light Gradient Boosting Machine Classifier (LightGBM) (in top 3 performances) and subsequently trained and evaluated it using Leave-One-Out (LOO) cross-validation. The model achieved a weighted F1 score of 0.73, with both precision and recall at 0.73.

We utilized SHAP (SHapley Additive exPlanations) \cite{lundberg2017unified} to examine which features were the most important to the model. Notably, the enrollment scores for role functioning (\textit{rl2}), cognitive functioning (\textit{cf}), RAI and both social and physical functioning were identified as the top five most important predictors of improvement. Gender or capsule prescription status did not play a role.

\begin{figure}
    \centering
    \includegraphics[width=.9\linewidth]{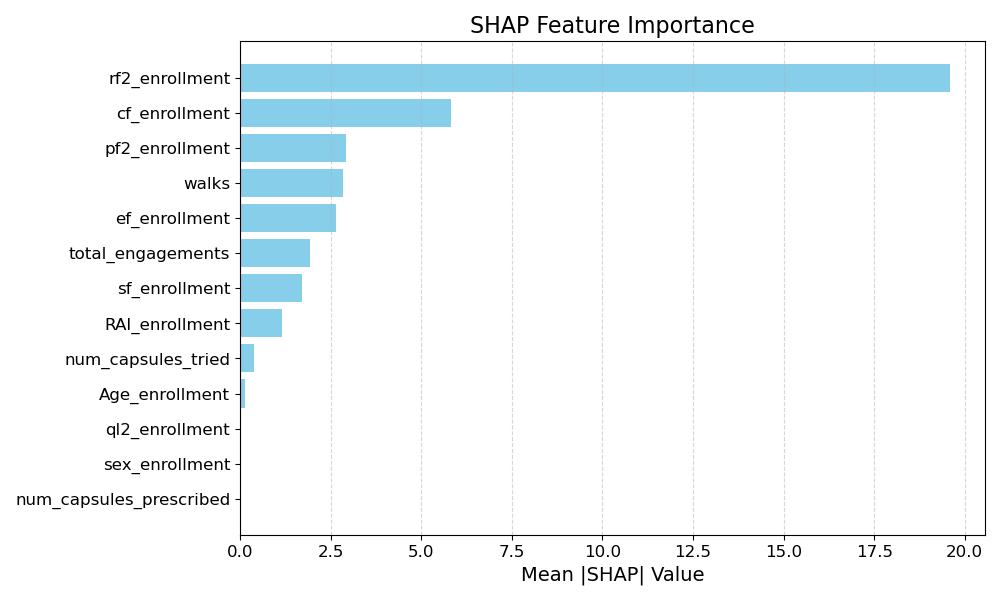}
    \caption{Feature Importance}
    \label{fig:fi}
\end{figure}

When we extended the analysis to incorporate micro-level engagement features such as the average number of weekly walks, the total number of capsules tried, and the total number of engagements with all capsules, the predictive F1 score, obtained from with LOO cross-validation, remained at 0.73 and the top 3 important feature remained the same (See Figure \ref{fig:fi}) . 

We analyzed the disparities between patients experiencing improvement in functioning and those showing no improvement or decline. Individuals who demonstrated improvement exhibited lower levels of emotional (p=0.02), social (p=0.02), role (p=0.002), and cognitive (p=0.004) functioning at the time of enrollment compared to those who did not, indicating room for improvement.  Among patients who completed the BREQ-2 questionnaire and showed improvement in functioning, RAI scores were higher than those who declined; however, this disparity is not statistical significance (p$>$0.05) (See Table \ref{tab:stat-diff}).

\section{Discussion}

In our study, cancer patients exhibited a high initial level of micro-level engagements, with many exploring multiple capsules independently. The walking capsule emerged as the most popular choice, likely influenced by motivational messages we sent to all patients encouraging them to walk. This preference aligns with our previous research, identifying walking as a favored behavior change intervention among both cancer patients and the healthy population, where the patients reported reminders have been beneficial for engagement \cite{lisowska2021personality, lisowska2022improve}.

We observed that sustained engagement with mobile DBCIs is notably enhanced by clinician prescriptions. This suggests that patients are more likely to persist with BCIs, such as walking, when recommended by their healthcare providers, underscoring the significant impact of clinician endorsement on the engagement with DBCIs.

Patients who managed to sustain or improve their physical or cognitive functioning engaged with a capsule on average once weekly, in contrast to those who showed a decline and did not engage after the first week. Thus, for cancer patients aiming to maintain well-being during treatment, \textit{effective engagement} appears to involve a minimum of one well-being activity weekly, although a causal link could not be established.  Yet when considering shifting motivation for engagement with well-being activity from extrinsic to intrinsic (i.e increasing RAI) to facilitate long term behaviour change the micro-level engagement might need to be closer to twice weekly. 

In predicting intervention outcomes, initial functioning was a key factor. Surprisingly, self-determined motivation for physical activity, captured by RAI at enrollment, did not directly contribute to outcome improvement but was associated with micro-level engagement.

Our findings did not reveal a direct link between micro-level engagement and quality of life improvements, despite lower engagement among patients with declining functioning. 
However, micro-level engagement was associated with a positive shift in self-determined motivation, which has yet to be linked to outcomes. The limited sample size presents one challenge, but another may be that changes in outcomes necessitate a long-term dedication to behavioral change. This process is incremental, beginning with micro-level interaction with the mobile application and gradually progressing through changes in the sources of motivation, potentially leading to improved well-being. A more extended period of follow-up could offer further insights. Additionally, the impact of pharmacological treatments on patients' function should be considered.

A significant limitation of our study was the absence of a control group, a consequence of our design, in which for ethical reasons, all patients received the same app and same features. 

Another challenge involved objectively measuring engagement with the intervention, since not all patients wore the provided watches. A few patients declined to use the ASUS watch, preferring their personal wearable devices instead. Therefore, a future application that allows patients to use their own watches and is compatible with various devices could more accurately capture actual engagement, particularly in interventions focused on walking.

\section{Summary Points}
\begin{itemize}
    \item The study aims to define effective DBCI engagement strategies for enhancing cancer patients' quality of life, using the CAPABLE project as a foundation.
    \item Metrics to measure engagement and hypotheses for DBCI impact assessment are developed, considering both patient and clinician interest in DBCIs 
    \item Findings highlight clinician-prescribed DBCIs significantly boost patient engagement.
    \item Regular, minimal use of DBCIs each week supports well-being; however, increased engagement is likely more effective in fostering a shift to intrinsic motivation for engaging in well-being activities.
\end{itemize}

\section{Acknowledgement}

The CAPABLE project has received funding from the European Union's Horizon 2020 research and innovation programme under grant agreement No 875052. 

 \bibliographystyle{elsarticle-num} 
 \bibliography{cas-refs}

\begin{thebibliography}{10}
\expandafter\ifx\csname url\endcsname\relax
  \def\url#1{\texttt{#1}}\fi
\expandafter\ifx\csname urlprefix\endcsname\relax\def\urlprefix{URL }\fi
\expandafter\ifx\csname href\endcsname\relax
  \def\href#1#2{#2} \def\path#1{#1}\fi

\bibitem{mummah2016ideas}
S.~A. Mummah, T.~N. Robinson, A.~C. King, C.~D. Gardner, S.~Sutton, Ideas (integrate, design, assess, and share): a framework and toolkit of strategies for the development of more effective digital interventions to change health behavior, Journal of Medical Internet Research 18~(12) (2016) e5927.

\bibitem{yardley2016understanding}
L.~Yardley, B.~J. Spring, H.~Riper, L.~G. Morrison, D.~H. Crane, K.~Curtis, G.~C. Merchant, F.~Naughton, A.~Blandford, Understanding and promoting effective engagement with digital behavior change interventions, American Journal of Preventive Medicine 51~(5) (2016) 833--842.

\bibitem{michie2017developing}
S.~Michie, L.~Yardley, R.~West, K.~Patrick, F.~Greaves, et~al., Developing and evaluating digital interventions to promote behavior change in health and health care: recommendations resulting from an international workshop, Journal of Medical Internet Research 19~(6) (2017) e7126.

\bibitem{parimbelli2021cancer}
E.~Parimbelli, M.~Gabetta, G.~Lanzola, F.~Polce, S.~Wilk, D.~Glasspool, A.~Kogan, R.~Leizer, V.~Gisko, N.~Veggiotti, et~al., {CAncer PAtients Better Life Experience (CAPABLE)} first proof-of-concept demonstration, in: International Conference on Artificial Intelligence in Medicine, Springer, 2021, pp. 298--303.

\bibitem{short2018measuring}
C.~E. Short, A.~DeSmet, C.~Woods, S.~L. Williams, C.~Maher, A.~Middelweerd, A.~M. M{\"u}ller, P.~A. Wark, C.~Vandelanotte, L.~Poppe, et~al., Measuring engagement in ehealth and mhealth behavior change interventions: viewpoint of methodologies, Journal of medical Internet research 20~(11) (2018) e292.

\bibitem{fabi2020cancer}
A.~Fabi, R.~Bhargava, S.~Fatigoni, M.~Guglielmo, M.~Horneber, F.~Roila, J.~Weis, K.~Jordan, C.~Ripamonti, Cancer-related fatigue: Esmo clinical practice guidelines for diagnosis and treatment, Annals of Oncology 31~(6) (2020) 713--723.

\bibitem{muradnew}
M.~Murad, N.~Asi, M.~Alsawas, F.~Alahdab, New evidence pyramid, BMJ Evidence-Based Medicine~(21) (2016) 125--127.

\bibitem{lisowska2023sato}
A.~Lisowska, S.~Wilk, M.~Peleg, Sato (ideas expanded with bcio): Workflow for designers of patient-centered mobile health behaviour change intervention applications, Journal of Biomedical Informatics 138 (2023) 104276.

\bibitem{aaronson1993european}
N.~K. Aaronson, S.~Ahmedzai, B.~Bergman, M.~Bullinger, A.~Cull, N.~J. Duez, A.~Filiberti, H.~Flechtner, S.~B. Fleishman, J.~C.~d. Haes, et~al., The european organization for research and treatment of cancer qlq-c30: a quality-of-life instrument for use in international clinical trials in oncology, JNCI: Journal of the National Cancer Institute 85~(5) (1993) 365--376.

\bibitem{lanzola2023case}
G.~Lanzola, F.~Polce, E.~Parimbelli, M.~Gabetta, R.~Cornet, R.~de~Groot, A.~Kogan, D.~Glasspool, S.~Wilk, S.~Quaglini, The case manager: An agent controlling the activation of knowledge sources in a fhir-based distributed reasoning environment, Applied Clinical Informatics 14~(04) (2023) 725--734.

\bibitem{agtabular}
N.~Erickson, J.~Mueller, A.~Shirkov, H.~Zhang, P.~Larroy, M.~Li, A.~Smola, Autogluon-tabular: Robust and accurate automl for structured data, arXiv preprint arXiv:2003.06505 (2020).

\bibitem{lundberg2017unified}
S.~M. Lundberg, S.-I. Lee, A unified approach to interpreting model predictions, Advances in neural information processing systems 30 (2017).

\bibitem{lisowska2021personality}
A.~Lisowska, S.~Lavy, S.~Wilk, M.~Peleg, Personality and habit formation: Is there a link?, in: Proceedings of the AIxIA 2021 SMARTERCARE Workshop, CEUR-WS, 2021, pp. 42--47.

\bibitem{lisowska2022improve}
A.~Lisowska, S.~Lavy, S.~Wilk, M.~Peleg, How to improve digital wellbeing interventions? preliminary study of factors affecting intervention engagement, impact, and habit formation, in: AMIA Annual Symposium Proceedings. American Medical Informatics Association, 2022.

\end{thebibliography}

\end{document}